\documentclass[twoside]{article}
\usepackage{fleqn,espcrc2}
\usepackage{graphicx}
\usepackage{here}

\newcommand{\AmS}{{\protect\the\textfont2
  A\kern-.1667em\lower.5ex\hbox{M}\kern-.125emS}}
\def\beq{\begin{equation}}
\def\eeq{\end{equation}}
\def\bea{\begin{eqnarray}}
\def\eea{\end{eqnarray}}
\def\bq{\begin{quote}}
\def\eq{\end{quote}}

\def\nnb{\nonumber}
\def\ga{\left(}
\def\dr{\right)}

\def\rar{\rightarrow}

\def\nnb{\nonumber}
\def\la{\langle}
\def\ra{\rangle}
\def\nin{\noindent}
\def\ba{\begin{array}}
\def\ea{\end{array}}
\def\bm{\overline{m}}

\def\als{\alpha_s}

\def\gg2{ \la\alpha_s G^2 \ra}
\def\gg3{g^3f_{abc}\la G^aG^bG^c \ra}
\def\ggg4{\la\als^2G^4\ra}

\title{\bf{\boldmath  
{\huge Muon and Tau Anomalies Updated} }}
\author{ 
Stephan Narison\address{ Laboratoire
de Physique Math\'ematique, Universit\'e
de Montpellier II Place Eug\`ene
Bataillon, 34095 - Montpellier Cedex 05,
France \\ and \\ Physics Division, National Center for Theoretical Sciences,
Hsinchu, Taiwan 300, Republic of China.\\ E-mail:
qcd@lpm.univ-montp2.fr} }
\begin{document}
\textwidth 17.6cm
\textheight 25.3cm
\topmargin -2.5cm
\oddsidemargin -.6cm
\evensidemargin -0.5cm
\pagestyle{empty}
\pagestyle{plain}
\begin{abstract}
\noindent
We present a new independent evaluation of the hadronic vacuum polarization contributions 
$a_l^{had}(l.o)\equiv \frac{1}{2}(g-2)_l^{had}(l.o)$ to the anomalous
magnetic moment ({\it anomaly}) of the muon  and tau leptons using $\tau$-decays and
$e^+e^-$ data. The alone theoretical input used for describing the
high-energy  region not accessible experimentally is perturbative QCD plus (negligible)
additional effects due to the QCD vacuum condensates. We obtain:
$a_\mu^{had}(l.o)= 7020.6(75.6)\times 10^{-11}$ and $a_\tau^{had}(l.o)= 353.6(3.8)\times
10^{-8}$, which we compare with previous determinations. Our analysis leads to the 
Standard
Model (SM) prediction:
$a_\mu^{SM}=116~591~846.9(78.9)\times 10^{-11}$.
Confronting $a_\mu^{SM}$ with the recent BNL measurement leads to
$a_\mu^{new}\equiv a_\mu^{exp}-a_\mu^{SM}=176(170)\times 10^{-11}$. Combined with 
the mean existing determinations of $a_\mu^{had}(l.o)$, this leads to the conservative range: $-42\leq 
a_\mu^{new}
\times 10^{11}\leq 413$ at 90\% CL, from which we derive lower bounds on the
scales of some new physics. 
We also update our old and first predictions of the SM contributions to
$a_\tau$. Including QED to sixth order, higher order hadronic and electroweak
contributions,  we obtain $a_\tau^{SM}=117~755(7)\times 10^{-8}$, waiting for a future
precise measurement of $a_\tau$.\\

\nin
March, 19th 2001\\
PM/01-13\\
NSC-NCTS-010314
\end{abstract}
\maketitle
\section{INTRODUCTION}
\nin
The recent E821 BNL result \cite{BNL99} for the measurement of the muon anomaly averaged
with older determinations \cite{CERN78,BNL98} gives \cite{MARCIANO}:
\beq\label{exp}
a_\mu^{exp}=116~ 592~ 023 (151)\times 10^{-11}~.
\eeq
The announcement of a $2.6\sigma$ deviation \cite{BNL99,MARCIANO} of this experimental
result from the standard model (SM) prediction has stimulated intensive publications on the
possible signal or/and constraints on new physics, 
but has also raised some criticisms
\cite{YND}. In this paper, we consider that the only rational answer to the
the criticisms of \cite{YND} is an independent re-evaluation of the
hadronic vacuum polarisation contribution to the muon anomaly as an update of our old estimate in
\cite{CALMET}, which is the main motivation of this work. In
order to avoid some specific theoretical dependences of the result, we shall mainly use the
available data from
$\tau$-decay and $e^+e^-\rar
$ hadrons, and limit ourselves to the use of perturbative QCD plus (negiligible) additional effects
due to the QCD vacuum condensates for describing the
high-energy region (QCD continuum) not accessible experimentally. We shall compare our results
with previous determinations and analyze its implications for
some new physics (e.g. supersymmetry, radiative muon, composite, extended Zee and leptoquark models) beyond
the SM. Finally, we update our old and first \cite{SNGM2} SM predictions for $a_\tau$.
\section{THE HADRONIC VACUUM POLARISATION CONTRIBUTION TO THE LEPTON ANOMALY}
\nin
Several papers have been devoted to the analysis of the contribution to the muon anomaly. Works prior 76 have
been reviewed in \cite{CALMET}, while more recent works (after 85) are reviewed in
\cite{MARCIANO,YND}. A partial historical review of the different determinations since 61 can be found in
Figure 2 of \cite{JEGER}, which we complete and update in Table 1, where only works published in
journals with referee-policies have been considered.  
\begin{table*}[hbt]
\setlength{\tabcolsep}{1.5pc}
\newlength{\digitwidth} \settowidth{\digitwidth}{\rm 0}
\catcode`?=\active \def?{\kern\digitwidth}
\caption{Time and precision evolutions of the determinations of $a_\mu^{had}(l.o)$ 
from $e^+e^-$ data. Results marked with * are our selected results which take into
account update and results with minimal theoretical inputs.}
\begin{tabular*}{\textwidth}{@{}l@{\extracolsep{\fill}}lll}
\hline
&\\
{$\bf a_\mu^{had}(l.o)\times 10^{11}$}&{\bf
~~~~~~~~~~~~~~~~~~~~~Authors}&{\bf Ref. }&{\bf Comments}\\ &&\\
\hline
&&\\
$\approx 3200$&Bouchiat-Michel (61)&\cite{BOUCHIAT}& $\pi^+\pi^-$ only\\ 
$5500\sim 11000$&Durand (62)&\cite{DURAND}&$\pi^+\pi^-$ only\\
$\approx 7500$&Kinoshita-Oakes (67)&\cite{KINO67}\\
$3400^{+1900}_{-900}$&Bowcock (68)&\cite{BOWCOCK}\\
6500(500)&Gourdin-de Rafael (69)&\cite{GOURDIN}\\
6800(900)&Bramon-Etim-Greco (72)&\cite{BRAMON}\\
7300(1000)&Bailey et al. (75)&\cite{BAILEY}\\
6630(850)&Barger-Long-Olsson (75)&\cite{BARGER}\\
6990(880)&Calmet-Narison-Perrottet-de Rafael (76\&77)&\cite{CALMET}\\
7020(800)*&Narison (78)&\cite{SNGM2}\\
6840(110)&Barkov et al. (85)&\cite{BARKOV}\\
7068(174)*&Kinoshita-Nizi\'c-Okamoto (85)&\cite{KINO85}\\
7100(116)*&Casa-Lopez-Yndurain (85)& \cite{CASAS}\\
7050(78)&Dubnicka-Martinovic (90)& \cite{DUB1}\\
7024(153)*&Eidelman-Jegerlehner (95) &\cite{EIDEL}\\
7113(103)&Adel-Yndurain (95)&\cite{ADEL}&\\
7026(160)*&Brown-Worstell (96)&\cite{BROWN}&\\
6950(150)*&Alemany-Davier-H\"ocker (97)&\cite{DAVIE}&\\
7011(94)&Alemany-Davier-H\"ocker (97) &\cite{DAVIE}&+$\tau$-decay\\
6951(75)*&Davier-H\"ocker (98) &\cite{HOCKERA}&+$\tau$-decay\\
6924(62)&Davier-H\"ocker (98) &\cite{HOCKER}&+Theoretical input\\
&&\\
7021(76)*&Narison (01)&This work& Average from $e^+e^-$ and\\
&&&$\tau$-decay (see section 3)\\
&&&\\
\hline
&&&\\
\multicolumn{3}{l}{\bf Informative weighted mean values}\\ 
&\\ 
\multicolumn{1}{l}{$6992(28)_{stat}$}
&& \multicolumn{2}{l}{Since 69}\\ 
\multicolumn{1}{l}{$7008(41)_{stat}$}
&& \multicolumn{2}{l}{Selected results marked with *}\\ 

 &&\\
\hline
\end{tabular*}
{\footnotesize 
\begin{quote}
\noindent
\end{quote}}
\end{table*}
\nin
Using a dispersion relation, the hadronic
vacuum polarisation contribution to the lepton anomaly $a_l$ can be expressed as
\cite{BOUCHIAT}--\cite{GOURDIN}:
\bea
a_l^{had}(l.o)=\frac{1}{4\pi^3}\int_{4m^2_\pi}^\infty dt~K_l(t)~\sigma_H(t)~.
\eea
$-$~$K_l(t\geq 0)$ is the QED kernel function \cite{LAUT}:
\bea\label{kernel}
K_l(t)&=&\int_0^1 dx\frac{x^2(1-x)}{x^2+\ga{t}/{m_l^2}\dr(1-x)}~,
\eea
with the analytic forms:
\bea
K_l(t\leq 4m_l^2)&=&\frac{1}{2}-4y_l-\nnb\\
&&4y_l(1-2y_l)\log(4y_l)\nnb\\
&&-2\ga 1-8y_l+8y_l^2\dr\times\nnb\\
&&\sqrt{y_l(1-y_l)}\arccos{\sqrt{y_l}}~,\nnb\\
K_l(t\geq 4m_l^2)&=&
z_l^2\ga 1-\frac{z_l^2}{2}\dr+
\ga
1+z_l\dr^2\times\nnb\\
&&\ga 1+
\frac{1}{z_l^2}\dr\Big{[}\log(1+z_l)-z_l\nnb\\
&&+\frac{z_l^2}{2}\Big{]}
+\ga\frac{1+z_l}{1-z_l}\dr z_l^2\log{z_l},
\eea
with:
\beq
y_l=\frac{t}{4m^2_l},~~z_l=\frac{(1-v_l)}{(1+v_l)}~~{\rm and
}~~v_l=\sqrt{1-\frac{4m_l^2}{t}}.
\eeq
 $K_l(t)$ is a monotonically decreasing function of $t$. For large $t$,
it behaves as:
\beq
K_l(t> m^2_l)\simeq \frac{m^2_l}{3t}~,
\eeq
which will be useful for the analysis in the large $t$ regime.
Such properties then emphasize the importance of the
low-energy contribution to $a_l^{had}(l.o)~(l\equiv e,~\mu)$, where the QCD analytic calculations cannot
be applied.\\
$-$~$\sigma_H(t)\equiv\sigma(e^+e^-\rar{\rm
hadrons})$ is the $e^+e^-\rar $ hadrons total cross-section which can be
related to the hadronic two-point spectral function Im$\Pi(t)_{em}$ through the
optical theorem:
\beq
R_{e^+e^-}\equiv\frac{\sigma(e^+e^-\rar{\rm
hadrons})}{\sigma(e^+e^-\rar\mu^+\mu^-)}=12\pi{\rm Im}\Pi(t)_{em}~,
\eeq
where: 
\beq
\sigma(e^+e^-\rar\mu^+\mu^-)=\frac{4\pi\alpha^2}{3t}.
\eeq
Here,
\bea\label{twopoint}
\Pi^{\mu\nu}_{em} &\equiv& i \int d^4x ~e^{iqx} \
\la 0\vert {\cal T}
{J^\mu_{em}(x)}
\ga {J^\nu_{em}(x)}\dr^\dagger \vert 0 \ra \nnb\\
&=&-\ga g^{\mu\nu}q^2-q^\mu q^\nu\dr\Pi_{em}(q^2)
\eea
is the correlator built from the local electromagnetic current:
\beq
J^\mu_{em}(x)=\frac{2}{3}\bar u\gamma^\mu u-
\frac{1}{3}\bar d\gamma^\mu d-\frac{1}{3}\bar s\gamma^\mu s+... 
\eeq
$-$~In the following,
we shall discuss in details the different hadronic contributions to $a_\mu^{had}(l.o)$. Our
results will be given in Table 2.
\section{HADRONIC VACUUM POLARISATION CONTRIBUTION TO THE MUON ANOMALY}
\subsection{LIGHT ISOVECTOR MESONS}
\subsection*{Region below 0.8 GeV$^2$}
\nin
Due to the $t$-behaviour of the kernel function $K_\mu(t)$, this region gives so far the
most important contribution to $a_\mu^{had}(l.o)$ ($\approx 68\%$ of the total contribution),
and also the largest source of the errors (77\% when added quadratically). In our numerical analysis, we divide
this region into three subregions.\\
\nin
$-$~
The first one is the region near the two pions threshold $4m_\pi^2\leq t \leq 0.4$
GeV$^2$, where the pion form factor $|F_\pi|^2(t)$ is constrained by universality
$|F_\pi|^2(0)=1$ and by the $t$ expansion predicted by chiral perturbation theory:
\beq
|F_\pi|^2(t)\simeq 1+\frac{1}{6}\la r^2\ra_\pi t+c_\pi t^2+{\cal O}(t^3)~,
\eeq
or its resummed expression \cite{PICH}. $\la r^2\ra_\pi\simeq (0.431\pm 0.026)~{\rm fm}^2$ is
the mean pion charge radius squared and $c_\pi\simeq (3.2 \pm 1.0)~{\rm GeV}^4$
\cite{COL}. One can
inspect that the ALEPH/OPAL \cite{ALEPH,OPAL} and $e^+e^-$ data compiled in
\cite{EIDEL,DAVIE,DOL} satisfy  both constraints.\\
\nin
$-$~ The second region is the one from $0.4 ~{\rm GeV}^2\leq t\leq 0.6 ~{\rm
GeV}^2$ on the top of the $\rho$ resonance.\\ 
$-$~ For these two regions, we shall use CVC hypothesis which relates the
electromagnetic to the charged current through an isospin rotation \cite{GILMAN}:
\beq
\sigma_H(t)=\frac{4\pi\alpha^2}{t}v_1~.
\eeq
We follow the notation of ALEPH \cite{ALEPH}, where:
\beq
{\rm Im}\Pi^{(1)}_{\bar ud,V}\equiv \frac{1}{2\pi}v_1~,
\eeq
is the charged vector two-point correlator:
\bea\label{twopoint}
\Pi^{\mu\nu}_{\bar ud,V} &\equiv& i \int d^4x ~e^{iqx} 
\la 0\vert {\cal T}
{J^\mu_{\bar ud}(x)}
\ga {J^\nu_{\bar ud}(0)}\dr^\dagger \vert 0 \ra \nnb\\
&=&-\ga g^{\mu\nu}q^2-q^\mu q^\nu\dr\Pi^{(1)}_{\bar ud,V}(q^2)\nnb\\
&&+q^\mu q^\nu\Pi^{(0)}_{\bar ud,V}(q^2)~,
\eea
built from the local charged current 
$
J^\mu_{\bar ud,V}(x)=\bar u\gamma^\mu d(x)~.
$
We use the
accurate semi-inclusive ALEPH/ OPAL data \cite{ALEPH,OPAL}. For a comparison, we also show the
results obtained from the use of $e^+e^-$ data in the whole region below 0.6 GeV$^2$.\\
\nin
$-$~ The third  region is the one from $0.6$ GeV$^2$ to $0.8$ GeV$^2$,
which is peculiar due to the $\omega-\rho$ mixing. In this region, we use either
the $\tau$-decay data with corrected $\omega-\rho$ mixing effect,
or the $e^+e^-$ data alone compiled in \cite{DAVIE,ALEPH}. One can check
from the data that the effect of the
$4\pi$ is negligible in such a region. 
\begin{table*}[hbt]
\setlength{\tabcolsep}{1.5pc}
\catcode`?=\active \def?{\kern\digitwidth}
\caption{Determinations of $a_l^{had}(l.o)$ using combined $e^+e^-$
and inclusive $\tau$ decay data (2nd and 4th columns) and averaged $e^+e^-$ data (3rd column).}
\begin{tabular*}{\textwidth}{@{}l@{\extracolsep{\fill}}llll}
\hline
&\\
\multicolumn{1}{l}{\bf Region in GeV$\bf ^2$}
 & \multicolumn{2}{c}{$\bf a_\mu^{had}(l.o)\times 10^{11}$}
  & \multicolumn{1}{l}{$\bf a_\tau^{had}(l.o)\times 10^{8}$}
& \multicolumn{1}{c}{\bf Data input} \\
&\\
\hline\\
                 & \multicolumn{1}{l}{\bf ${\bf\tau}$+$\bf e^+e^-$}
& \multicolumn{1}{l}{$\bf e^+e^-$} 
                 & \multicolumn{1}{l}{\bf ${\bf\tau}$+$\bf e^+e^-$}\\ 
{\bf Light Isovector}&&&\\
$4m_\pi^2\rar 0.8$&$4794.6\pm 60.7$&$4730.2\pm 99.9$&$165.8\pm 1.5$&\cite{DAVIE,ALEPH,OPAL}\\
$0.8\rar 2.1$&$494.9\pm 15.8$&$565.0\pm 54.0$&$28.7\pm 1.1$&\cite{ALEPH,OPAL}\\
$2.1\rar 3.$&$202.0\pm 29.7$&$175.9\pm 16.0$&$17.0\pm 2.6$&\cite{ALEPH,OPAL}\\
\it Total Light I=1&$\it 5491.5\pm 69.4$&$\it 5471.1\pm 114.7$&$\it 211.5\pm 3.2$\\
\bf Light Isoscalar&&\\
{\it Below 1.93 }&\\
$\omega$&$387.5\pm 13$&$387.5\pm 13$&$15.3\pm 0.5$&NWA \cite{PDG}\\
$\phi$&$393.3\pm 9.9$&$393.3\pm 9.9$&$21.0\pm 0.5$&NWA \cite{PDG}\\
$0.66\rar 1.93$&$79.3\pm 18.7$&$79.3\pm 18.7$&$4.3\pm 1.1$&$\sum{\rm exclusive}$ \cite{DOL}\\
{\it From 1.93 to 3~} &\\
$\omega(1.42),~\omega(1.65)$&$31.3\pm 6.8$&$31.3\pm 6.8$&$2.6\pm 0.7$&BW \cite{DM2,PDG}\\
$\phi(1.68)$&$42.4\pm 18.2$&$42.4\pm 18.2$&$3.8\pm 1.3$&BW \cite{DM2,DM1,PDG}\\
{\it Total Light I=0 }&$\it 933.8\pm 31.5$ &$\it 933.8\pm 31.5$&$\it 47.0\pm 2.0$&\\
\bf Heavy Isoscalar&&\\
$J/\psi(1S\rar 4.415)$&$87.0\pm 4.7$&$87.0\pm 4.7$&$13.08\pm 0.69$&NWA \cite{PDG}\\
$\Upsilon(1S\rar 11.020)$&$0.95\pm 0.04$&$0.95\pm 0.04$&$0.23\pm 0.01$&NWA \cite{PDG}\\
\it Total Heavy I=0&$\it 88.0\pm 4.7$&$\it 88.0\pm 4.7$&$\it 13.3\pm 0.7$\\
\bf QCD continuum&&\\
$3.\rar (4.57)^2$&$407.0\pm 2.3$&$407.0\pm 2.3$&$49.4\pm 0.3$&$(u,d,s)$ \\
$(4.57)^2\rar (11.27)^2$&$95.3\pm 0.5$&$95.3\pm 0.5$&$27.3\pm 0.1$&$(u,d,s,c)$ \\
$(11.27)^2\rar 4M^2_t$&$20.5\pm 0.1$&$20.5\pm 0.1$&$5.87\pm 0.01$&$(u,d,s,c,b)$ \\
$4M^2_t\rar \infty$&$\approx 0.$&$\approx 0.$&$\approx 0.$&$(u,d,s,c,b,t)$\\
\it Total QCD Cont.&$\it 522.8\pm 2.4$&$\it 522.8\pm 2.4$&$\it 82.6\pm 0.3$&\\
&\\
\hline
&\\
&7036.1(76.4)&7015.7(119.1)&354.4(3.8)&\\
&\\
\hline\\
\end{tabular*}
\end{table*}
\nin
\subsection*{Region from 0.8 to 3 GeV$^2$}
\nin
For our fitting procedure, we use CVC and the ALEPH/OPAL semi-inclusive $\tau$-decay data
\cite{ALEPH,OPAL}. The dominant error from this region comes from the one between 2.1 and 3
GeV$^2$ (88\% when the errors are added quadratically), due to the incaccuracy of the
data near $M_\tau$. For a comparison, we also
show the results when using the sum of exclusive modes from $e^+e^-$ compiled in \cite{ALEPH,DAVIE},
where in this case the errors come mainly from the region below 2 GeV$^2$.
\subsection{LIGHT ISOSCALAR MESONS}
\subsection*{$\omega$ and $\phi$ mesons}
\nin
We treat these mesons in a narrow width approximation (NWA), which is expected to be a
good approximation. Using the
relation between the $e^+e^-\rar$ hadrons total cross section and the leptonic width
$\Gamma_{ee}$ of the meson with a mass $M_R$:
\beq
\sigma_H(t)_{NWA}\simeq 12\pi^2\frac{\Gamma_{ee}}{M_R}\delta(t-M^2_R)~,
\eeq
one can write:
\beq
a_\mu^{had}(l.o)_{NWA}\simeq\frac{3}{\pi}\frac{\Gamma_{ee}}{M_R}K_\mu(M^2_R)~,
\eeq
where, we shall use the PDG values of the electronic widths \cite{PDG}.
\subsection*{Region below $1.39^2$ GeV$^2$}
\nin
We estimate the effect of this region by using the sum of exclusive $I=0$ modes
for the ratio $R^{I=0}_{e^+e^-}$ compiled in \cite{DOL}.
\subsection*{Region between $1.39^2$ and 3 GeV$^2$}
\nin
In order to account for the multi-odd pions, $\bar KK,~\bar KK\pi,...$ modes, we estimate the effect
of this region by assuming that it is mainly given by the
$\omega(1419), ~\omega(1662)$ and $\phi(1680)$, with their parameters measured
by the DM2 \cite{DM2} and DM1 \cite{DM1} collaborations. We shall estimate their 
leptonic widths which are multiplied by the hadronic
branching ratios in \cite{DM2,DM1}, by assuming (see PDG
\cite{PDG} and  the indication from DM2 and DM1 data), that the $\omega(1419)$ decays predominantly into
$\rho\pi$ (we assume it to be approximately 90\%); the 
$\omega(1662)$ decays mainly into $\rho\pi$ (44\%) and $\omega\pi$ (56\%), while the
$\phi(1680)$ decays dominantly into $K^*K$ (93\%) and to $K\bar K$ (7\%). In this way,
we deduce the leptonic widths:
\bea
\Gamma_{\omega(1419)\rar e^+e^-}&\simeq& (90\pm 34)~{\rm eV}~ ,\nnb\\
\Gamma_{\omega(1650)\rar e^+e^-}&\simeq& (0.30\pm 0.03)~{\rm keV}~ ,\nnb\\
\Gamma_{\phi(1680)\rar e^+e^-}&\simeq& (0.43\pm .15)~{\rm keV}~.
\eea
We use a Breit-Wigner (BW) form for evaluating their contributions:
\beq
\sigma_H(t)\vert_{BW}\simeq 12\pi\frac{\Gamma_{ee}\Gamma_{tot}}
{(t-M^2_R)^2+M^2_R\Gamma^2_{tot}}~,
\eeq
where we take $M_R$ from PDG \cite{PDG} and:
\bea
\Gamma_{tot}^{\omega(1420)}&\simeq& (174\pm 59)~{\rm MeV}~\cite{DM2},\nnb\\
\Gamma_{tot}^{\omega(1650)}&\simeq& (280\pm 24)~{\rm MeV}~\cite{DM2},\nnb\\
\Gamma_{tot}^{\phi(1680)}&\simeq& (150\pm 50)~{\rm MeV}~\cite{PDG}.
\eea
\subsection{THE $J/\psi$ AND $\Upsilon$ FAMILIES}
\nin
We consider the six $J/\psi$ mesons $1S$, $2S,$ 3770, 4040,
4160 and 4415, and the five $\Upsilon$ mesons $1S$, $2S$, $3S,$ $4S$
and 11020. We treat these mesons using NWA and the data on their electronic 
widths from PDG \cite{PDG}.
\subsection{QCD CONTINUUM}
\nin
As advertized previously, we shall treat the region not accessible experimentally
by using perturbative QCD where it is expected to work and where non-perturbative
effects like e.g. the quark and gluon condensates are negligible corrections. 
\subsection*{Input for light quarks}

In the case of massless quarks, the unit operator contribution to $R_{e^+e^-}$ is accurately known and
reads, in the
$\overline{MS}$ scheme:
\bea
R_{e^+e^-}^{pert}= 3\sum_f Q^2_f\big{[} 1+a_s+F_3a_s^2+F_4a_s^3\big{]}
\eea
with:
\bea\label{massless}
F_3&=&1.986-0.115n_f~~~\cite{ALFA2}~,\nnb\\
F_4&=&-6.637-1.200n_f-0.005n^2_f\nnb\\
&&-1.240\frac{\ga{\sum_f Q_f}\dr^2}{3\ga\sum_f Q^2_f\dr}~~~\cite{ALFA3}~,
\eea
for $n_f$ quark flavours; $Q_f$ is the quark charge in units of $e$
and $a_s\equiv
(\alpha_s/\pi)$ is the QCD running coupling which we use to order $a_s^2$ (see e.g.
the expression in \cite{BNP,SNB}). We also use, the value of the QCD scale for three flavours:
\bea\label{inputL}
\Lambda_3&=&(375\pm 50)~{\rm MeV},
\eea
obtained using the central value $\alpha_s(M_Z)=0.119$ \cite{PDG,BETHKE}.

To this perturbative correction,
we add the light quark mass and non-perturbative corrections. We shall also
consider the effects of a tachyonic gluon mass evaluated in \cite{CNZ} in order
to take into account the effects of the truncation of the QCD perturbative series.

The dimension $D=2$ ($m_f^2$ and tachyonic gluon) corrections are \cite{BNP,CNZ,CHET2}:
\bea
tR_{e^+e^-}^{(2)}&=&-3\sum_f Q^2_f\bigg{[}{\bar m^2_f}\big{[} 6+28a_s+\nnb\\
&&(294.8-12.3n_f)a_s^2\big{]}+1.05a_s\lambda^2
\bigg{]}~,
\eea
where \cite{CNZ} $a_s\lambda^2\simeq -(0.06\sim 0.07)$ GeV$^2$. 

The dimension $D=4$
corrections are \cite{SVZ,BNP}:
\bea
t^2R_{e^+e^-}^{(4)}&=&3\sum_f Q^2_f
\Bigg{[}\frac{2\pi}{3}\ga 1-\frac{11}{18}a_s\dr\la \alpha_s G^2\ra\nnb\\
&&+
8\pi^2\ga 1-a_s\dr m_f\la \bar \psi_f\psi_f\ra\nnb\\
&&+\frac{32\pi^2}{27}a_s\sum_fm_f\la \bar \psi_f\psi_f\ra\nnb\\
&&+\bar{m}_s^4\frac{8}{7}\Big{[}-\frac{6}{a_s}+\frac{23}{4}\nnb\\
&&+a_s\ga
\frac{2063}{24}-10\zeta_3\dr\Big{]}\Bigg{]}~,
\eea
where $\zeta_3=1.2020569$...

We use $\la \alpha_s G^2\ra=(0.07\pm 0.01)$ GeV$^4$  \cite{SNG},
and $(m_u+m_d)\la\bar uu+\bar dd\ra=-2f_\pi^2m_\pi^2$ ($f_\pi=92.6$ MeV).
The running quark mass in the $\overline{MS}$ scheme is defined as (see e.g. \cite{SNB}):
\bea
\bm_i(\nu)&=&\hat{m}_i\ga -\beta_1 a_s(\nu)\dr^{-\gamma_1/\beta_1}
\Bigg\{1\nnb\\&&+\frac{\beta_2}{\beta_1}\ga \frac{\gamma_1}{\beta_1}-
 \frac{\gamma_2}{\beta_2}\dr a_s(\nu)~
\nnb\\&&+\frac{1}{2}\Bigg{[}\frac{\beta_2^2}{\beta_1^2}\ga \frac{\gamma_1}
{\beta_1}-
 \frac{\gamma_2}{\beta_2}\dr^2-
\frac{\beta_2^2}{\beta_1^2}\ga \frac{\gamma_1}{\beta_1}-
 \frac{\gamma_2}{\beta_2}\dr\nnb\\&&+
\frac{\beta_3}{\beta_1}\ga \frac{\gamma_1}{\beta_1}-
 \frac{\gamma_3}{\beta_3}\dr\Bigg{]} a^2_s(\nu)\Bigg\}~,
\eea
where for three flavours the quark mass anomalous dimensions are (see e.g. \cite{BNP,SNB}):
$\gamma_1=2$, $\gamma_2=91/12,$ $\gamma_3=24.8404$, while the coefficients of the
$\beta$ function are: $\beta_1=-9/2,$ $\beta_2=-8,$ $\beta_3=-20.1198 $.
We shall use the value of the invariant strange quark mass \cite{SNL}:
\beq\label{inputl} 
\hat{m}_s=(133.3\pm 18.8)~{\rm MeV}~.
\eeq 
We take into account the $SU(3)$
breaking of the quark condensates $\la\bar ss\ra/
\la\bar dd\ra\simeq 0.68^{+0.15}_{-0.29}$ \cite{SNC} and consistently use $f_K=1.2f_\pi$.
\subsection*{Input for heavy quarks}
\nin 
In the
case of heavy quarks, the perturbative spectral function can be accurately
approximated by the Schwinger interpolating formula \cite{SCHWIN}:
\beq\label{massive}
R_{e^+e^-}=3\sum_f Q^2_f~ v_f\frac{(3-v_f^2)}{2}\big{[} 1+\frac{4}{3}\alpha_s f(v_f)\big{]}~,
\eeq
where:
\beq
f(v_f)=\frac{\pi}{2v_f}-\frac{(3+v_f)}{4}\ga\frac{\pi}{2}-\frac{3}{4\pi}\dr~,
\eeq
and:
\beq
v_f=\sqrt{1-\frac{4m_f^2}{t}}~,
\eeq
is the quark velocity; $m_f$ is the heavy quark (pole) mass of flavour $f$. We shall see that, in
the region where we shall work (away from threshold and for large $t$), these two
parametrizations  provide a sufficiently accurate description of the spectral
function. For a conservative estimate, we shall consider the range of
values spanned by the running and pole quark masses for the heavy quarks \cite{SNH}:
\bea\label{inputh}
m_c=(1.2\sim 1.46)~{\rm GeV},~m_b=(4.2\sim 4.7)~{\rm GeV}.
\eea
Within the present accuracy of the values of the heavy
quark masses, we find that it is not necessary to include the known
 $\alpha_s^2$ corrections \cite{CHET}.
\subsection*{The $(u,d,s)$ flavour contributions}
\nin
 We complete the contribution of the light quark channels by adding the QCD
continuum from $t_c=3$ to (4.57 GeV)$^2$.
We shall use the value of $\Lambda_3$ given in Eq. (\ref{inputL}). 
Our value of $t_c$ is higher than $t_c\simeq 1.6$ GeV$^2$ from global duality
constraint in the $I=1$ channel \cite{PERIS}\footnote{I thank
Eduardo de Rafael for this remark. A similar value of the QCD continuum is obtained from the
$V$-$~A$ channel \cite{SNCP}.}. We have
compared $a_\mu^{I=1}(1.6\leq t\leq 3)$ GeV$^2$ from a QCD continuum parametrization
and from the uses of data. They only differ by 16$\times 10^{-11}$, showing that
the QCD continuum provides a good smearing of the data. We shall include
this value into the systematics.
One can also realize that the mass and non-perturbative corrections tend
to cancel each others, whilst, individually, they are already small.
\subsection*{The $ (u,d,s,c)$ flavour contributions}
\nin
We add the contribution of the charm quark about 150 MeV above the
$J/\psi(4415)$ ($\approx$ empirical mass-splitting between the
radial excitations). In this region, the previous expressions give a good
description of the spectral function, because at this energy, the charm quark is already
relativistic with a velocity
larger than 0.75. Additional threshold effects are taking into account by
transforming the value of
$\Lambda_3$ into $\Lambda_4$. 
\subsection*{The $ (u,d,s,c,b)$ flavour contributions}
\nin
The contribution of the $b$-quark is added about 250 MeV above the
$\Upsilon$(11020) resonance ($\approx$ empirical mass-splitting between the
radial excitations). Again at this energy, the $b$ quark is already relativistic as its velocity
is larger than 0.55, such that our perturbative parametrization remains a good approximation. We integrate
until the
$2M_t$ threshold, where we  take $M_t\simeq 174.3(5.1)$ GeV as a best value given by PDG \cite{PDG}.
\subsection{FINAL RESULT FOR $a_\mu^{had}(l.o)$}
\subsection*{From combined $\tau$-decay and $e^+e^-$ data}
\nin
Collecting all different contributions from Table 2, we deduce the final result
at the end of that table, which we 
compile in Table 1 for a comparison with other previous determinations. We obtain
\footnote{Here and in the following, the errors will be added quadratically as usually done.}: 
\beq\label{rtauf}
a_\mu^{had}(l.o)=7036.1(76.4)\times 10^{-11}~,
\eeq
using the inclusive $\tau$-decay data until 3 GeV$^2$ from ALEPH/OPAL \cite{ALEPH,OPAL} for the
isovector channel and using $e^+e^-$ data below 3 GeV$^2$ \cite{PDG,DOL,DM2,DM1} for the isoscalar
channel. The errors have been added quadratically. The effects of the heavy quark mesons have been
treated using a narrow width approximation plus QCD continuum away from the quark-anti-quark thresholds.
\subsection*{From the alone $e^+e^-$ data}
\nin
If we use the isovector $e^+e^-$ data below 3 GeV$^2$, we obtain:
\beq\label{eplus}
a_\mu^{had}(l.o)=7015.7(119.1)\times 10^{-11}~,
\eeq
which is slightly lower than the one from inclusive $\tau$ decay and less accurate. The difference in each
region can be seen explicitly in Table 2 and easy to understand from 
the data given in \cite{ALEPH,DAVIE}.
\subsection*{Different source of errors}
\nin
$-$~ The
main error (80\% when added quadratically) in our previous determinations comes from the $\rho$-meson
region below 0.8 GeV$^2$. Hopefully, improved
measurements of this region are feasible in the near future. \\
\nin
$-$~ The second source of errors 
comes from the region around
$M_\tau$ for the inclusive $\tau$-decay and between 1 GeV to $M_{\tau}$  for the $e^+e^-$ data. These
errors are about half of the one from the region below 0.8 GeV$^2$ in most different determinations.
They can be reduced by improved measurements of inclusive $\tau$-decay near $M_\tau$ $(I=1)$ and by
improving the measurements of the odd multi-pions and $\bar KK, ~\bar KK\pi,...$ channels in
the $I=0$ channels from $e^+e^-$ data.\\ 
$-$~The
contributions of the whole region above 3 GeV$^2$ induce much smaller errors (7\% of the total).
There is a quite good consensus between different determinations in this energy region. 
\subsection*{Improvements from  $\tau$-decay data}
\nin
Our results for the isovector contributions, show that the precise
inclusive
$\tau$-decay data have significantly improved by almost a factor two the accuracy of the
determination compared with the one from $e^+e^-$ data. However, they are amended by the 
inaccuracy near $M_\tau$. Indeed, one can gain a bit in precision by using $e^+e^-$ data in the region
between 2.1 and 3 GeV$^2$ instead of $\tau$-decay data, but we are aware of the disagreement
among different $e^+e^-$ data in this region. In this case of figure,
one would get from Table 2:
\beq\label{hybrid}
a_\mu^{had}(l.o)=7010.0(72.2)\times 10^{-11}~.
\eeq
\subsection*{Final result and additional systematic errors}
\nin
We consider as
{\it a final result} the average of the three results given in Eqs.
(\ref{rtauf}), (\ref{eplus}) and (\ref{hybrid}). We consider
the accuracy given in Eq. (\ref{hybrid}), which avoids the weakness of
$\tau$-decay data near $M_\tau$. We add to this error the following systematics:
\\
$-$~ We take into account, an
eventual deviation from the CVC assumption \footnote{I thank Paco
Yndurain for mentioning this point, for checking some of the entries in Table 2 and for several
communications.} (CVC has been also tested in \cite{PICHSN}), and some other systematics in
manipulating the isovector data by adding a systematic error of 15.5$\times 10^{-11}$ coming from
the largest difference of the central value of Eqs. (\ref{rtauf}) and (\ref{eplus}) from the
average value. \\ 
$-$~ We also add another 16$\times 10^{-11}$ systematics due to the difference of the
continuum threshold 3 GeV$^2$ value for three flavours used here and the one from QCD global
duality in
\cite{PERIS} (see section 3.4).
\\
Adding the different errors quadratically, we deduce {\it our final estimate}:
\beq\label{final}
a_\mu^{had}(l.o)=7020.6(75.6)\times 10^{-11}~,
\eeq
It is amusing to notice that the central value of our different results almost co\"\i ncide
with our 25 years old value \cite{CALMET,SNGM2} and with the {\it mean central value} given in Table~1.
\subsection{COMPARISONS WITH PREVIOUS DETERMINATIONS OF $a_\mu^{had}(l.o)$}
\subsection*{QCD continuum}
\nin
$-$~ Our treatment of the QCD continuum in the heavy quark channels gives similar results than
more involved parametrization \cite{ADEL} including non-perturbative gluon condensate and
non-relativistic contributions. This is due to the fact that in the region where we work,
the heavy quark-antiquark pairs are already relativistic with a quark velocity larger than 0.55.\\
\nin
$-$~
Comparing existing estimates, one can see that there is a consensus on the size of the QCD
continuum effect. The departures from different determinations come mainly from the
treatment of the low-energy region below 2 GeV.
\subsection*{Isoscalars}
\nin
For the isoscalar resonances, one can realize by comparing the results with
existing estimates that the narrow width approximation (NWA) gives enough accurate
results comparable in magnitude and accuracy with more sophisticated parametrization 
including width effects \cite{EIDEL}.
\subsection*{Isovector channel and some global comparisons}
\nin
In the following comparison, we shall use numbers in units of $10^{-11}$.\\
$-$~{\it Comparison with \cite{CASAS,ADEL}}\\
It has been argued in \cite{CASAS} that 
the contribution from this region can be improved by using constraint
imposed by $\pi\pi$ scattering data in the spacelike region on the pion form factor.
The result $a_\mu^{had}(l.o)(t\leq 0.8 $ GeV$^2)=4848(31)$ quoted by \cite{CASAS} is based on
Watson theorem (extended to 0.8 GeV$^2$) for the $\pi\pi$ phase shift $\delta^1_1(t)$.
The central value differs with our value 4795(61) obtained from the
data in the timelike region by 53 MeV, but agrees with the
result 4794(14) given in their Eq. (3.23) obtained from a fit of $\delta^1_1(t)$ from the data. Another
discrepancy of 39(104) comes from the region between 0.8 and 2 GeV$^2$, but is not significant
due to the large errors. It is argued in \cite{ADEL} that error from
high-energy region can be reduced .
\\
\nin
$-$~{\it Comparison with \cite{EIDEL}}\\
Our result agrees within the errors
with the final result 7024(153) of \cite{EIDEL} from their Table 3a \footnote{I thank William
Marciano for mentioning this correct result and for several communications.}, with the
updated result 6989(111) quoted in
\cite{JEGER}, and with some of the results given in Table 2.
\\
$-$~{\it Comparison with \cite{DAVIE,HOCKERA,HOCKER}}

~ For a proper comparison with these different determinations, we shall only consider the two
regions separated by the scale $\sqrt t=M_\tau$, and compare our result $a_\mu^{had}(l.o)(\sqrt t\leq
M_\tau)= 6426(76)$ obtained from $\tau$-decay data. The central value of our result is
relatively higher than the value 6365(74) in \cite{HOCKERA} and 6343(60) in \cite{HOCKER}. Within
the errors, the 1st result is consistent with ours while the second shows one $\sigma$ discrepancy.
For comparing with the result of \cite{DAVIE}, we subtract from their
total result, the heavy quarks and QCD continuum contributions. In this way, we deduce the
contribution 6405(94), much more closer to our result than the two formers, where one should also
notice that  our final results from the alone $e^+e^-$ data agree.

~ The exact origin of the discrepancy of the central values is not easy to detect due to the
complexity of the analysis in this region, though it is reassuring that results based on
maximal data inputs \cite{DAVIE,HOCKERA} are consistent with ours within the errors. 

~ We do not worry by the discrepancy with
\cite{HOCKER}, as it uses
more theoretical inputs than data (``determination with minimal data input"
as quoted from \cite{SCHILCHER}) for minimizing the errors , through the
introduction (and subtraction) of some arbitrary polynomials in $t$ (not coming from QCD first principles)
for  approaching the
$t=0$ region~\footnote{This procedure together with the analytic
continuation method 
has been already criticized in
\cite{SNB}. Analogous comments apply to the estimate of $\alpha(M_Z)$.}. On the other hand, the use of ``local
duality" in a given interval is stronger than the  standard applications of QCD spectral sum rules \`a la SVZ
\cite{SVZ} (for a review see e.g.
\cite{SNB}), where systematic uncertainties can be more severe. We do not
expect that the result obtained in this way is much better (though apparently more accurate) than the ones
previously obtained from pure data analysis.

~ One should also notice
that, though one has been able to formulate the $\tau$-decay width using QCD and the Operator Product
Expansion (OPE) including the QCD vacuum condensates
\cite{BNP,LEDI} \`a la SVZ \cite{SVZ}, which has provided an accurate measurement of the QCD
coupling $\alpha_s$ \cite{ALEPH,OPAL}, the extension of such a program to 
the analysis of the lepton anomaly cannot be done in a straightforward way due the existence of a pole
at $t=0$, which needs a nonperturbative QCD control of the contribution of the small circle at the
origin when using the Cauchy contour for evaluating the muon anomaly in the complex-t plane. Such a
control is not feasible at present from QCD first principles.
\subsection*{Final errors and comments on the uses of new data}
\nin 
We can conclude from the previous
determinations in Table 1 and the present work that, with present data and with less theoretical
inputs, a realistic error for extracting $a_\mu^{had}(l.o)$ cannot be less than $70\times
10^{-11}$. After the completion of this work, 
we have checked \cite{SNALFA} by doing
similar analyses (hadronic contributions to $\alpha$ and to the muonium hyperfine splitting) that,
using the new Novosibirsk
\cite{NOVO} and BES
\cite{BES} data, one does not improve noticeably the errors compared to the
one obtained from the correlated averaged data and QCD continuum parametrizations used in this
paper.
\section{THEORY OF THE MUON ANOMALY}
\subsection{SUM OF THE THEORETICAL CONTRIBUTIONS}
\nin
$-$~{\it QED and SM electroweak contributions}\\
Using obvious notations, they read \cite{MARCIANO,HUGHES,YND}:
\bea\label{qed}
a_\mu^{QED}&=&116~ 584~ 705.7(2.9)\times 10^{-11}~,\nnb\\
a_\mu^{EW}&=&151(4)\times 10^{-11}~.
\eea
with QED up to 8th order and EW including two-loop corrections \cite{ZARN}.\\
$-$~{\it Higher order hadronic contributions}\\
For $a_\mu^{had}(h.o)$, we have used the contribution of the
high-order vacuum polarizations
$(h.o)_{V.P}$ recently estimated in \cite{KRAUSE}, where we have checked that the central value co\"\i
ncide with the original estimate in \cite{CALMET} which has the same result for
$a_\mu^{had}(l.o)$ though the error is
larger in \cite{CALMET} due to older data. This feature is reassuring for a
self-consistency check. As the data used in
\cite{CALMET,KRAUSE} give smaller value of $a_\mu^{had}(l.o)$ by 1.00125 than the present work, we have
rescaled the
$(h.o)_{V.P}$ contribution by this small factor, and we deduce:
\beq
a_\mu^{had}(h.o)_{V.P}=-101.2(6.1)~.
\eeq
For the light by light scattering hadronic contributions,
 we use the average of the two estimates in \cite{KINO2} which agree each others. 
We have corrected the sign of the pseudoscalar and
axial-vector contributions according to the
result in \cite{KNECHT}, which has been confirmed in \cite{KINO01} \footnote{We are also
aware of the fact that the two results are model-dependent and further tests on their
validity are needed. However, one can notice that this contribution is of the same size
as the error of
$a_\mu^{had}(l.o)$.}. Then, we obtain:
\beq\label{lbl}
a_\mu^{had}(h.o)_{LL}=70.8(21.0)\times 10^{-11}~.\eeq
We have not included into the sum the contribution due to the quark loop, suspecting that
adding this effect together with the pion loop and meson pole contributions can lead
to a double counting (quark-hadron duality arguments \footnote{The agreement of the result
in Eq. (\ref{lbl}) with the one $62(3)\times 10^{-11}$ obtained in \cite{KINO85} (see also \cite{TROCO}),
using the quark constituent model (and no hadrons at all!) seems to support such arguments.}). Instead, we
have considered the quark loop effect incorporating vector mesons in the photon legs, which is about
$15\times 10^{-11}$ (average of the two determinations in
\cite{KINO2}) as another source of uncerainties of the estimate. Then, we deduce:
\bea\label{hadronic}
a_\mu^{had}(h.o)&\equiv& a_\mu^{had}(h.o)_{V.P}+a_\mu^{had}(h.o)_{LL}\nnb\\
&=&-30.4(21.9)\times 10^{-11}~,
\eea
showing that the higher order hadronic effects tend to cancel each others.\\
$-$~{\it Total theoretical predictions in the SM}\\
Adding the results in Eqs. (\ref{qed}) and (\ref{hadronic}), one obtains 
the SM theoretical contributions:
\bea
a_\mu^{SM}&=&116~ 584~ 826.3(22.5)\times 10^{-11}+a_\mu^{had}(l.o)\nnb\\
&=&116~591~846.9(78.9)\times
10^{-11}~,
\eea
where $a_\mu^{had}(l.o)$ is the lowest order hadronic contributions evaluated in this paper
\big{[}see Eq. (\ref{final})\big{]}.
\subsection{THEORY VERSUS EXPERIMENT}
\nin
In confronting the theoretical estimate with the experimental
value in Eq. (\ref{exp}), we shall use, for reasons explained previously, the result in Eq. (\ref{final}) with
a moderate accuracy. We deduce:
\beq\label{range}
a_\mu^{new}\equiv a^{exp}_\mu-a^{SM}_\mu=176(170)\times 10^{-11}~.
\eeq
The mean central value from our selected determinations given in Table 1 would give:
\beq\label{range1}
\la a_\mu^{new}\ra\equiv a^{exp}_\mu-\la a^{SM}_\mu\ra =189(175)\times
10^{-11}~,
\eeq
after adding to the ``statistical" error in Table 1, the one from our determination in Eq.
(\ref{final}).  Both values indicate an {\it eventual window} of about only $\sigma$ for {\it
new physics} beyond the standard model. A more definite claim is waiting for a more precise
$a_\mu$ measurement. We translate the results in Eqs. (\ref{range}) and (\ref{range1}) into {\it
the~conservative~ range}: 
\beq\label{rangef}
-42\leq a_\mu^{new}\times 10^{11}\leq 413~~~({\rm 90\% ~CL})~.
\eeq
We shall discuss some numerical implications of this result  by giving conservative lower bounds
on the scales of new physics  discussed recently in the literature \cite{MARCIANO,MARCIA2,LANE,NG,KING}. 
\subsection{LOWER BOUNDS ON NEW PHYSICS}
\subsection*{Supersymmetry}
\nin
Using the expression given in \cite{MARCIANO} and for $\tan\beta
\equiv \la \phi_2\ra/\la \phi_1\ra$ (ratio of
Higgs expectation values) $\geq 4$:
\beq
a_\mu^{SUSY}\simeq \frac{\alpha}{8\pi\sin^2\theta_W}\frac{m^2_\mu}{\tilde m^2}\tan\beta~,  
\eeq
 one obtains a 
conservative lower bound on the degenerate sparticle mass:
\beq
\tilde m\ge 113~{\rm GeV}~~( 90\%~{\rm CL}),
\eeq
which is comparable with the present experimental lower bound of 100 GeV. 
\subsection*{Radiative mass, composite and extended Zee models}
\nin
In these models where the typical contribution to $a_\mu$ is
 about $ ({m^2_\mu}/{M^2}){\cal O}(1) ~
$ \cite{MARCIANO,MARCIA2,LANE},
 one can derive the lower bound on the scale of the models:
\beq
M\ge 1.7~{\rm TeV}~.
\eeq
This constraint can also be translated into a lower bound on the mass of a singlet scalar
in the Zee Model with a  coupling $1/96\pi^2$ \cite{NG}.
One can deduce:
\beq
M_{S_2}\ge 55~{\rm GeV}~.
\eeq
\subsection*{Leptoquarks scenarios}
\nin
Leptoquark effects on $a_\mu$ have been discussed recently in \cite{KING}.
For an electromagnetic coupling at the lepton-leptoquark-quark vertex, one can deduce
the approximate expression:
as:
\beq
a_\mu^{lq}\simeq\ga
\frac{\alpha}{\pi}\dr\frac{m_cm_\mu}{M^2_{lq}}\log{\frac{M^2_{lq}}{m_c^2}}~,
\eeq
giving:
\beq
M_{lq}\ge 1.1~{\rm TeV}~.
\eeq
This bound is much larger than the present lower bounds of about $(200\sim 300)$~GeV from direct search
experiments at HERA and Tevatron. \\
We expect that these different bounds will
be improved in the near future both from accurate measurements of $a_\mu$ and of $e^+e^-$ data necessary
for reducing the theoretical errors in the determinations of the hadronic contributions, being the major
source of the theoretical uncertainties.
\section{THEORY OF THE TAU ANOMALY}
\nin
Here, we update our {\it old and first work} \cite{SNGM2} on the $\tau$ lepton anomaly. We
shall use $M_\tau=1.77703$ GeV from PDG \cite{PDG},
\subsection*{Lowest order hadronic contribution}
\nin 
The estimate from $e^+e^-+\tau$-decay is given in Table 2. Taking the similar average as in the
case of the muon (section 3.5), one obtains:
\beq\label{tauhadlo}
a_\tau^{had}(l.o)=353.6(4.0)\times 10^{-8}~.
\eeq
One can see that the estimate $280(20)\times
10^{-8}$ in \cite{NASR}, using present values of the QCD condensates \cite{SNG,SNB}, is much lower than this
result. This appears to be a general feature of
this method based on minimal data inputs (see e.g. \cite{HOCKER} for $a_\mu^{had}(l.o)$
discussed in the previous section). A comparison with the result of
\cite{EIDEL} shows that in both cases of muon and tau, this result  is systematically higher than
ours by the same scaling factor 1.0004. Previous results 370(40) \cite{SNGM2} and 360(30)
\cite{MENDEL} in units of
$10^{-8}$ are comparable with ours but inaccurate.
\subsection*{Higher order hadronic contributions}
\nin
The higher order contributions due to vacuum polarization $a_\tau^{had}(h.o)_{V.P}$, can be obtained from the
result of
\cite{KRAUSE}. We  rescale it by the factor 1/1.0083 like we have done for the muon case as explained in
previous section 9. For the light by light scattering contribution, we use the result of
\cite{KINO2} for the muon and we rescale it by the mass squared ratio $(m_\tau/m_\mu)^2$, which is
expected to be a good approximation from the semi-analytical expression given in \cite{CALMET}. Then, we
obtain:
\bea
a_\tau^{had}(h.o)_{V.P}&=&7.6(0.2)\times 10^{-8}~,\nnb\\
a_\tau^{had}(h.o)_{L.L}&=&\ga\frac{m_\tau}{m_\mu}\dr^2 a_\mu^{had}(h.o)_{L.L}\nnb\\
&=&20.0(5.8)\times 10^{-8}~.
\eea
Therefore, we deduce:
\beq\label{tauhadho}
a_\tau^{had}(h.o)=27.6(5.8)\times 10^{-8},
\eeq
and:
\beq
a_\tau^{had}\equiv a_\tau^{had}(l.o)+
a_\tau^{had}(h.o)=381.2(7.0)\times 10^{-8}.
\eeq
\subsection*{Electroweak contributions}
\nin
The lowest order electroweak SM 
contribution is well-known:
\bea
a_\tau^{EW}(l.o)= \frac{5}{3}\frac{G_\tau
m^2_\tau}{8\pi^2\sqrt{2}}\Big{[}1+\frac{1}{5}(1-4\sin^2\theta_W)^2
\Big{]},
\eea
to order $(m_\tau/M)^2$ ($M$ being the W or Higgs mass); We assume $G_\tau =G_\mu=
1.166~39(1)\times 10^{-5}$ is the Fermi coupling; $\sin^2\theta_W=0.224$, where $\theta_W$ is the weak
mixing angle. We add the two-loop contribution
(2-loop/l.o$\simeq -65\alpha/\pi)$ \cite{ZARN}, which induces a 15\% reduction of the one
loop result. We obtain:
\beq
a_\tau^{EW}\simeq 46.9(1.2)\times 10^{-8}~.  
\eeq
\subsection*{QED contributions}
\nin
$-$~{\it Generalities}\\
These contributions have been first evaluated in \cite{SNGM2} to
order
$\alpha^3$. Here, we revise and improve these
evaluations. In so doing, we use the relation \cite{PETER,KINO3}:
\bea
a_\tau-a_e&=&\sum_{l,l'\equiv
e,\mu}\Big{[}a_4{(l)}\ga\frac{\alpha}{\pi}\dr^2+\nnb\\
&&\big{[}
a_6{(l)}+
a_6{(ll')}\big{]}\ga\frac{\alpha}{\pi}\dr^3\Big{]},
\eea 
where $l$ indicates the internal fermion loop appearing in the photon propagator. $a_4{(l)}$ and $
a_6{(l)}$ can be deduced from the result of the muon anomaly while $a_6{(ll')}$ is a new contribution
involving one electron and one muon loop insertion in one photon propagator. At the same order of
truncation of PT series, one has \cite{KINO3,HUGHES}:
\bea
a_e&=& 0.5\ga\frac{\alpha}{\pi}\dr-0.328~478~965\ga\frac{\alpha}{\pi}\dr^2+\nnb\\
&&1.181~241~456\ga\frac{\alpha}{\pi}\dr^3+...\nnb\\
&=&1~159~65.2\times 10^{-8}~,
\eea
where we have used the measurement:
\beq
\alpha^{-1}=137.036~003~7(33)~,
\eeq 
from the quantum Hall effect \cite{HALL}.\\
$-$~{\it QED at fourth order}\\
For evaluating this contribution,
we use the dispersive representation:
\bea
a_4{(l)}=\frac{1}{3}\int_{4m^2_l}^\infty \frac{dt}{t^2}
(t+2m_l^2)\sqrt{1-\frac{4m_l^2}{t}}~K_\tau(t)~,
\eea
where $K_\tau(t)$ is the kernel function defined in Eq. (\ref{kernel}).
Then, we obtain:
\beq
a_4{(e)}=2.024~29~~~~~{\rm and}~~~~~a_4{(\mu)}=0.361~66~.
\eeq
The former result is well approximated by the known analytic approximate relation to order
$(m_e/m_\tau)^2$ given e.g. in
\cite{CALMET}, while the second result needs the inclusion of the not yet available $(m_\mu/m_\tau)^3$
term \footnote{These results agree with the last paper in \cite{MENDEL} confirming
that the results in their two former papers are wrong.}.
Adding these two contributions, we obtain:
\beq
\ga a_\tau-a_e\dr_4=2.385~95\ga\frac{\alpha}{\pi}\dr^2=1287.3\times 10^{-8}
\eeq
$-$~{\it QED at sixth order}\\
Because of the accurate determinations of the hadronic and weak contributions, the inclusion of
the sixth order contributions becomes necessary as they contribute with the same strength.
The contributions of diagrams with vacuum polarizations and ladders, can
be obtained from the analytic result
given for the muon in \cite{KINO3}:
\bea\label{a6l}
a_6(l)&=&\frac{2}{9}\log^2{\frac{m_\tau}{m_l}}-1.113~90\log{\frac{m_\tau}{m_l}}+\nnb\\
&&4.307~66+...~,
\eea
which leads to:
\beq
a_6(e)=10.000~2~,~~~~~~~~~~a_6(\mu)=2.934~0~.
\eeq
$a_6(\mu)$ differs from the value $-.122$ given in \cite{MENDEL} where the origin of
the negative value is not understandable from Eq. (\ref{a6l}) but should come
from a wrong term used there. We add the new class of contributions 
specific for $a_\tau$ given in \cite{SNGM2}:
\bea
a_6(e\mu)&=&2\Bigg{\{}\frac{2}{9}\log{\frac{m_\tau}{m_e}}\log{\frac{m_\tau}{m_\mu}}-
\frac{25}{54}\log{\frac{m^2_\tau}{m_e m_\mu}}\nnb\\
&&+\frac{\pi^2}{27}+\frac{317}{324}\Bigg{\}}=2.753~16~.
\eea
We parametrize the light by light scattering contribution as:
\beq
a_6(l)_{L.L}\simeq \frac{2\pi^2}{3}\log{\frac{m_\tau}{m_l}}-14.13~,
\eeq
by combining the known coefficient of the log-term \cite{SAMUEL} and the numerical value of the total
contribution for the muon \cite{KINO4}.
This leads to:
\beq\label{a6light}
a_6(e)_{L.L}=39.521~7~,~~~~~~~~~~a_6(\mu)_{L.L}=4.441~2~,
\eeq
 This contribution is so far the most important at
sixth order. Adding the different contributions from Eqs (\ref{a6l}) to (\ref{a6light}), one
obtains to sixth order:
\beq
\ga a_\tau-a_e\dr_6=59.650\ga\frac{\alpha}{\pi}\dr^3=74.758\times 10^{-8}~.
\eeq
This effect is about the same strength as the weak interaction effect and bigger than the higher order
hadronic contributions.\\
$-$~{\it Total QED contribution up to sixth order}\\
Adding the previous QED contributions, we deduce
\beq
a_\tau^{QED}=117~327.1(1.2)\times 10^{-8}~,
\eeq
where the error comes from the measurement of $\alpha$.
\subsection*{Final result}
\nin
Summing up the previous different theoretical contributions, we deduce in the standard model:
\bea\label{atau}
a_\tau^{SM}&=&a_\tau^{had}+a_\tau^{EW}+a_\tau^{QED}\nnb\\
&=&117~755.2(7.2)\times 10^{-8}~.
\eea
We consider this result as an improvement of our old \cite{SNGM2} and other
existing results. In \cite{MARCIA2}, only
an average of different existing hadronic contributions have been added to the
QED and electroweak contributions. In \cite{SNGM2,MENDEL}, the hadronic contributions
are inaccurate, while the value of the sixth order QED contribution is incorrect. 
This value in Eq. (\ref{atau}) can be compared with the present (inaccurate) experimental value
\cite{TAYLOR}:
\beq
a_\tau^{exp}=0.004\pm 0.027\pm 0.023~,
\eeq
which, we wish, will be improved in the near future.
\section{SUMMARY}
\nin
$-$~We have re-evaluated the lowest order hadronic contribution to the muon and tau anomalies
using the
precise $\tau$-decay data below $M_\tau$ and averaged $e^+e^-$ data. Ours results are given in Table 2
and section 7, and our best choice in Eq. (\ref{final}). Though
the approach is not conceptually new, the present situation of $a_\mu$ (theory versus experiment) justifies
a new independent re-evaluation of the anomaly using minimal theoretical inputs. \\
\nin
$-$~ We have extensively discussed the different sources
of the errors in the analysis, which are dominated by the region below $M_\tau$. However, the
relative weight of this region decreases for $a_\tau$ (see also detailed discussions in
 \cite{SNGM2}) where a more
precise theoretical prediction can then be provided. \\
\nin
$-$~We have compared our analysis summarized in Table 2 and our final result in Eq.
(\ref{final}) with some of the existing estimates given in Table 1
\footnote{During the submission of this paper to the journal, some other papers appeared in the
literature
\cite{YND01}.}. \\
\nin
$-$~We have shortly discussed implications of our result to some models beyond the standard model,
namely supersymmetry with large $\tan\beta$, radiative muon mass, composite, some extended Zee and
leptoquark models. The lower bounds on the mass scale of these models from the muon anomaly is
comparable or in some cases stronger than existing experimental lower bounds. \\
$-$~We have completed our work by updating our old estimate \cite{SNGM2} of the 
different theoretical contributions (QED up to sixth order, higher order hadronic  and electroweak) to the
$\tau$ lepton anomaly. Our result is given in Eq. (\ref{atau}). A precise future measurement of the
$\tau$-lepton  anomaly is welcome as it will permit to probe the QED series at shorter distance, and where,
the relative weights of different interaction effects are very different from the case of the muon. 
\section*{ACKNOWLEDGEMENTS}
\nin
It is a pleasure to thank Hsiang-nan Li and the theory
group of NCTS-Hsinchu for the warm hospitality.


\begin{thebibliography}{999}
\bibitem{BNL99}H.N. Brown et al., hep-ex/0102017.
\bibitem{CERN78}J. Bailey et al., {\it Nucl. Phys.} {\bf B 150} (1979) 1.
\bibitem{BNL98}H.N. Brown et al., hep-ex/0009029.
\bibitem{MARCIANO}A. Czarnecki and W.J. Marciano, hep-ph/0102122.
\bibitem{YND}F.J. Yndurain, hep-ph/0102312.
\bibitem{CALMET} 
J. Calmet, S. Narison, M. Perrottet and E. de Rafael, 
{\it Rev. Mod. Phys.} {\bf 49} (1977) 21; {\it Phys. Lett.} {\bf B 161 } (1976) 283;
S. Narison, {\it Th\`ese de 3\`eme cycle}, Marseille (1976).
\bibitem{SNGM2}S. Narison, {\it J. Phys. (Nucl. Phys.)} {\bf G 4} (1978) 1840.
\bibitem{JEGER}F. Jegerlehner, 
hep-ph/0104304.  
\bibitem{BOUCHIAT}C. Bouchiat and L. Michel, J. Phys. Radium {\bf 22} (1961) 121.
\bibitem{DURAND}L. Durand III, {\it Phys. Rev.} {\bf 128} (1962) 441; erratum {\bf 129} (1963) 2835.
\bibitem{KINO67}T. Kinoshita and R. Oakes, {\it Phys. Lett.} {\bf B 25} (1967) 143.
\bibitem{BOWCOCK}J.E. Bowcock, {\it Z. Phys.} {\bf 211} (1968) 400.
\bibitem{GOURDIN}M. Gourdin and E. de Rafael, {\it Nucl. Phys.} {\bf B 10} (1969) 667.
\bibitem{BRAMON}A. Bramon, E. Etim and M. Greco, {\it Phys. Lett.} {\bf B 39} (1972) 514.
\bibitem{BAILEY}J. Bailey et al., {\it Phys. Lett.} {\bf B 55} (1975) 420.
\bibitem{BARGER}V. Barger, W.F. Long and M.G. Olsson, {\it Phys. Lett.} {\bf B 60} (1975) 89.
\bibitem{BARKOV}L.M. Barkov et al., {\it Nucl. Phys.} {\bf B 256} (1985) 365.
\bibitem{KINO85}T. Kinoshita, B. Nizi\'c and Y. Okamoto, {\it Phys. Rev.} {\bf D 31} (1985) 2108.
\bibitem{CASAS}J.A. Casas, C. L\'opez and F.J. Yndur\'ain, {\it Phys. Rev.} {\bf D 32} (1985) 736.
\bibitem{DUB1}S. Dubnicka and L. Martinovic, {\it Phys. Rev.} {\bf D 42} (1990) 884.
\bibitem{EIDEL}S. Eidelman and F. Jegerlehner, {\it Z. Phys.} {\bf C 67} (1995) 585, and private
communication.
\bibitem{ADEL} K. Adel and F.J. Yndurain, hep-ph/9509378 (1995), {\it Rev. Acad. Ciencias (Esp)},
{\bf 92} (1998). 
\bibitem{BROWN}D.H. Brown and W.A. Worstell, {\it Phys. Rev.} {\bf D 54} (1996) 3237. 
\bibitem{DAVIE} R. Alemany, M. Davier and A. H\"ocker, {\it Eur. Phys.~J.}
{\bf C 2} (1998) 123.
\bibitem{HOCKERA}M. Davier and A. H\"ocker, {\it Phys. Lett.} {\bf B 419}
(1998) 419.
\bibitem{HOCKER}M. Davier and A. H\"ocker, {\it Phys. Lett.} {\bf B 435}
(1998) 427.
\bibitem{LAUT}B. Lautrup and E. de Rafael, {\it Phys. Rev.} {\bf 174} (1968) 1835.
\bibitem{PICH} D. G\'omez Dumm, A. Pich and J. Portol\'es, hep-ph/000332.
\bibitem{COL}G. Colangelo, M. Finkemeir and R. Urech, {\it Phys. Rev.} {\bf D 54} (1996) 4403.
\bibitem{ALEPH}The ALEPH collaboration, R. Barate et al., 
{\it Eur. Phys. J.} {\bf C 76} (1997) 15; {\bf C 4} (1998) 409; A. Hocker,
hep-ex/9703004 (I thank Michel Davier for
bringing the first reference to my attention).
\bibitem{OPAL} The OPAL collaboration, K. Ackerstaff et al., {\it Eur. Phys. J.} {\bf C 7} (1999)
571.
\bibitem{GILMAN}F.J. Gilman and S.H. Rhie, {\it Phys. Rev.} {\bf D 31} (1985) 1066.
\bibitem{PDG}PDG 2000, D.E. Groom et al., {\it Eur.
Phys. J.} {\bf C 15} (2000) 1.
\bibitem{DOL}S. Dolinsky et al., {\it Phys. Rep.} {\bf C 202} (1991) 99.
\bibitem{DM2}The DM2 collaboration, A. Antonelli et al., {\it Z. Phys.} {\bf C 56} (1992) 15;
D. Bisello et al., {\it Z. Phys.} {\bf C 39} (1988) 13.
\bibitem{DM1}The DM1 collaboration, F. Mane et al., {\it Phys. Lett.} {\bf B 112} (1982) 178;
A. Cordier et al., {\it Phys. Lett.} {\bf B 110} (1982) 335.
\bibitem{ALFA2}K.G. Chetyrkin, A.L. Kataev and F.V. Tkachov, {\it Phys. Lett.} {\bf B
85} (1979) 277 ;
M. Dine and J. Sapirstein, {\it Phys. Rev. Lett.} {\bf 43} (1979) 668 ;
W. Celmaster and R. Gonsalves, {\it Phys. Rev. Lett.} {\bf 44} (1980) 560.
\bibitem{ALFA3} S.G. Gorishny, A.L. Kataev and S.A. Larin,
{\it Phys. Lett.} {\bf B 259} (1991) 144;
See also : L.R. Surguladze and M.A. Samuel, {\it Phys. Rev. Lett.}
{\bf 66} (1991) 560 and erratum 2416.
\bibitem{BNP}E. Braaten, S. Narison and A. Pich, {\it Nucl. Phys.} {\bf B 373} (1992) 581.
\bibitem{SNB}S. Narison, {\it Phys. Rep.} {\bf 84} (1982) 263; {\it QCD spectral sum
rules, Lecture Notes in Physics,} {\bf Vol 26} (1989) 19, ed. World Scientific;
{\it The QCD theory of hadrons}, Cambridge Monograph Series (to appear in 2001).
\bibitem{BETHKE}S. Bethke, 
{\it Nucl. Phys. (Proc. Suppl.)} {\bf B, C 39} (1995); 
ibid, {\bf B, A 54} (1997); hep-ex/0004201;
 M. Schmelling, {ICHEP96}, Varsaw,1996;
I. Hinchliffe and A. Manohar hep-ph/0004186.
\bibitem{CNZ}K.G. Chetyrkin, S. Narison and V.I. Zakharov, {\it Nucl. Phys.} {\bf B 550} (1999) 353.
\bibitem{CHET2}K.G. Chetyrkin and J.H. K\"uhn, {\it Phys. Lett.} {\bf B
248} (1990) 359.
\bibitem{SVZ}M.A. Shifman, A.I. Vainshtein and V.I. Zakharov, 
{\it Nucl. Phys}~{\bf B 147} (1979) 385, 448.
\bibitem{SNG}S. Narison, {\it Phys. Lett.} {\bf B 387} (1996) 162; ibid {\bf
B 361} (1995) 121 and references therein.
\bibitem{SNL}S. Narison, {\it Nucl. Phys. (Proc. Suppl.)} {\bf B 86} (2000) 242; 
{\it Phys. Lett.} {\bf B 216} (1989) 191; H. Leutwyler, hep-ph/001049.
\bibitem{SNC}S. Narison, {\it Phys. Lett.} {\bf B 358} (1995) 113.
\bibitem{SCHWIN}J. Schwinger, {\it Particles, Sources and Fields}, Vol 2, Addison-Wesley (1973).
\bibitem{SNH}S. Narison, {\it Phys. Lett.} {\bf B 341} (1994) 73.
\bibitem{CHET}K.G. Chetyrkin, J.H. K\"uhn and M. Steinhauser, {\it Nucl. Phys.} {\bf B 482} (1996)
213.
\bibitem{PERIS}S. Peris, M. Perrottet and E. de Rafael, {\it Journ. High Ener. Phys.} {\bf 05} (1998)
01.
\bibitem{SNCP}S. Narison, {\it Nucl. Phys.} {\bf B 593} (2001) 3; {\it Nucl. Phys. (Proc. Suppl.)} {\bf B 96}
(2001) 364.
\bibitem{PICHSN}S. Narison and A. Pich, {\it Phys. Lett.} {\bf B 304} (1993) 359.
\bibitem{SCHILCHER}K. Schilcher, {\it in Conf. on Non-perturbative methods, Montpellier} (1985) World
Scientific, ed. S. Narison and references therein; S. Groote et al., {\it Phys. Lett.} {\bf B 440} (1998)
375.
\bibitem{LEDI}F. Le Diberder and A. Pich, {\it Phys. Lett.} {\bf B 286} (1992) 147; {\bf B 289} (1992) 165.
\bibitem{SNALFA}S. Narison, hep-ph/0108065.
\bibitem{NOVO}The CMD collaboration, R.R. Akhmetshin et al., {\it Nucl. Phys.} {\bf A 675} (2000) 424c; S.I.
Serednyakov,  {\it Nucl. Phys. (Proc. Suppl.} {\bf B 96} (2001) 197.
\bibitem{BES}The BES collaboration, J.Z. Bai et al., {\it Phys. Rev. Lett.} {\bf 84} (2000) 594;
hep-ex/0102003.
\bibitem{HUGHES}V.W. Hughes and T. Kinoshita, {\it Rev. Mod. Phys.} {\bf 71, 2} (1999)
S133.
\bibitem{ZARN}A. Czarnecki, B. Krause and W.J. Marciano, {\it Phys. Rev. Lett.} {\bf 76} (1996) 3267.
\bibitem{KRAUSE}B. Krause, {\it Phys. Lett.} {\bf B 390} (1997) 392.
\bibitem{KINO2}M. Hayakawa and T. Kinoshita and A. Sanda, {\it Phys. Rev.} {\bf D 54} (1996) 3137;
M. Hayakawa and T. Kinoshita, {\it Phys. Rev.} {\bf D 57} (1998) 465; J. Bijnens, E. Pallante and J.
Prades, {\it Nucl. Phys.} {\bf B 474} (1996) 379 and private communication from J. Prades; E. de
Rafael, {\it Phys. Lett.} {\bf B 322} (1994) 239.
\bibitem{KNECHT}M. Knecht and A. Nyffeler, hep-ph/0111058;\\ M. Knecht, A. Nyffeler,
M. Perrottet and E. de Rafael, hep-ph/0111059.
\bibitem{KINO01}M. Hayakawa and T. Kinoshita, hep-ph/0112102;\\ I. Blokland, A. Czarnecki
and K. Melnikov, hep-ph/0112117.
\bibitem{TROCO}J.F. de Troc\'oniz and F.J. Yndurain, hep-ph/0106025.
\bibitem{MARCIA2}A. Czarnecki and W.J. Marciano, {\it Nucl. Phys. (Proc. Suppl.)} {\bf B 76} (1999) 245.
\bibitem{LANE}E. Eichten, K. Lane and J. Preskill, 
{\it Phys. Rev. Lett.} {\bf 47} (1980) 225; K. Lane, hep-ph/0102131 and private communication; X. Calmet, H.
Fritzsch  and D. Holtmannsp\"otter, hep-ph/0103012.
\bibitem{NG}D.A. Dicus, H-J. He and J.N. Ng, hep-ph/0103126 and private communication from J.N. Ng. 
\bibitem{KING} K. Cheung, hep-ph/0102238 and references therein, and private communication; D.
Chakraverty, D. Choudhury and A. Datta, hep-ph/0102180.
\bibitem{NASR}F. Hamzeh and N.F. Nasrallah, {\it Phys. Lett.} {\bf B 373} (1996) 211.
\bibitem{MENDEL}G. Li, R. Mendel and M.A. Samuel, {\it Phys. Rev. Lett.} {\bf 67} (1991) 668;
erratum {\bf 69} (1992) 995; {\it Phys. Rev.} {\bf D 47} (1993) 1723.
\bibitem{PETER}B.E. Lautrup, A. Peterman and E. de Rafael, {\it Phys. Rep.} {\bf C 3} (1972) 193.
\bibitem{KINO3}W.J. Marciano and T. Kinoshita, {\it in Quantum Electrodynamics, Advanced Series
on Directions in High Energy Physics},{\bf Vol 7}, World Scientific., ed. Kinoshita.
\bibitem{HALL}E. Kr\"uger, W. Nistler and W. Weirauch, {\it Metrologia} {\bf 32} (1995) 117.
\bibitem{SAMUEL}B.E. Lautrup and M.A. Samuel, {\it Phys. Lett.} {\bf B 72} (1977) 114.
\bibitem{KINO4}T. Kinoshita, {\it Phys. Rev. Lett.} {\bf 61} (1988) 2898.
\bibitem{TAYLOR}L. Taylor, {\it Nucl. Phys. (Proc. Suppl.)} {\bf B 76} (1999) 273.
\bibitem{YND01}W.J. Marciano and B. Lee Roberts, hep-ph/0105056; J.F. de Troc\'oniz and F.J. Yndurain,
hep-ph/0106025; K. Melnikov, hep-ph/0105267; E.
Barto\v{s} et al., hep-ph/0106084.
\end{thebibliography}
\end{document}